\RequirePackage{fix-cm}
\documentclass[smallextended]{svjour3}       
\smartqed  
\usepackage{appendix}
\begin{document}

\title{A semiclassical condition for chaos based on Pesin theorem
}


\author{I. Gomez \and M. Losada \and S. Fortin \and M. Castagnino \and M. Portesi}


\institute{I. Gomez \at Instituto de F\'{i}sica de Rosario (IFIR-CONICET) and Depto. de F\'{i}sica (UNLP) \\
              Tel.: +54 (0341) 4853200 / 4853222, int: 422 \\
              \email{stratoignacio@hotmail.com} \\             
             \emph{Present address: BV. 27 de Febrero 210 Bis
Rosario,	 Santa Fe,	 2000,	 Argentina.} 
\and M. Losada \at Instituto de F\'{i}sica de Rosario (IFIR-CONICET) \\
Tel.: +54 (0341) 4853200 / 4853222, int: 422 \\ 
\email{marcelolosada@yahoo.com} \\
\emph{Present address: BV. 27 de Febrero 210 Bis
Rosario,	 Santa Fe,	 2000,	 Argentina.} 
\and S. Fortin \at  Depto. de F\'{i}sica, FCEN (UBA-CONICET) \\
Tel.: (5411) 4576-3300 \\
\email{sebastian.fortin@gmail.com} \\
\emph{Present address: Ciudad Universitaria - Pabell\'{o}n II - Ciudad Aut\'{o}noma de Buenos Aires, Argentina.} 
\and M. Castagnino \at Instituto de F\'{i}sica de Rosario (IFIR-CONICET) \\
Tel.: +54 (0341) 4853200 / 4853222, int: 486 \\ 
\email{castagninomario@gmail.com}\\
 \emph{Present address: BV. 27 de Febrero 210 Bis
Rosario,	 Santa Fe,	 2000,	 Argentina}
\and M. Portesi \at Instituto de F\'{i}sica de La Plata (IFLP-CONICET) and Depto. de F\'{i}sica (UNLP) \\
Tel.: (54-221) 4247201-4246062– 4230122, int: 245 \\
\email{portesi@fisica.unlp.edu.ar}\\
\emph{Present address: C.C. 67, 1900 La Plata, Argentina.}} 

\date{Received: date / Accepted: date}

\maketitle

\begin{abstract}
A semiclassical method to determine if the classical limit of a quantum system is chaotic or not, based on Pesin theorem, is presented. The method is applied to a phenomenological Gamow--type model and it is concluded that its classical limit is chaotic.
\keywords{Pesin theorem \and Lyapunov exponents \and Kolmogorov--Sinai entropy \and classical limit}
\end{abstract}

\section{Introduction}
\label{intro}

The presence of Lyapunov exponents in quantum systems has been reported in several papers \cite{z1,z2,z3,z4,JP,Pas} and the positiveness of them is a necessary condition for chaos. In classically chaotic quantum systems, decoherence formalism can be used to define quantum chaos. In such case the purity exponentially decreases at a Lyapunov rate \cite{paz2,paz3}.

A complete definition of classical chaos can be found in
\cite{BE}, where the three most important features of chaos, Lyapunov
exponents, Ergodic Hierarchy (EH) and complexity are studied. Brudno
theorem is the link between Kolmogorov--Sinai (KS) entropy and complexity, while Pesin theorem is the link between Lyapunov exponents and
KS--entropy as defined in EH \cite{3M}.

In this paper we focus on the Pesin theorem, which states that KS--entropy of the system, i.e. the average unpredictability of information of all possible trajectories in the phase space, is equal to the sum of all positive Lyapunov exponents.

A reasonable definition of quantum systems with a chaotic classical description has been given by M. Berry: ``\emph{A quantum system is chaotic if its classical limit is chaotic}"\cite{BERRY1}. This \emph{quantum caology}, as has been named originally by Berry, is what later came to be called quantum chaos.

In previous works \cite{0,NACHOSKY MARIO} some of us studied the quantum ergodic hierarchy (QEH). It ranks the chaotic level of quantum systems according to how quantum correlations between states and observables are
canceled for large times. From mixing level of QEH we can define a classical statistical limit which allows to reconcile chaos with the Correspondence Principle \cite{NACHOSKY MARIO2}. In \cite{NACHOSKY MARIO} we used QEH to characterize typical chaos phenomena, like the exponential localization of kicked rotator and the quantum interference destruction of Casati--Prosen model in terms of ergodic and mixing levels.
Moreover, QEH is an attempt, among several theoretical and phenomenological approaches (like WKB approximation or random matrix theory \cite{stockmann,haake,gutzwiller,casati,tabor}), to define a framework for quantum chaos which admits a chaotic classical description assuming Berry's definition.

In this paper we use QEH idea of ranking quantum chaos with quantum mean values, to present a semiclassical condition for chaos by means of Pesin theorem. More precisely, we express classical quantities by means of quantum mean values, using the Wigner transformation. In particular, we apply this technique to Pesin theorem.

The paper is organized as follows. In Section 2 we present the KS--entropy and Pesin theorem. In Section 3 we review the Wigner transformation, that we employ in the next sections. In Section 4 we express the Pesin theorem by quantum mean values and we obtain a semiclassical condition for chaos that gives a method to determine chaos in the classical limit. In Section 5 we apply this method to a phenomenological Gamow model \cite{Omnes1,PRE} and we conclude that its classical limit is chaotic. Finally, in Section 6 we discuss and draw some conclusions.

\section{Kolmogorov--Sinai entropy and Pesin theorem}
\label{sec:1}

We give the general notions of KS--entropy and Pesin theorem within the standard framework of measure theory.
We consider a dynamical system $(\Gamma, \Sigma, \mu, \{T_t\}_{t\in J})$, where $\Gamma$ is the phase space, $\Sigma$ is a $\sigma$-algebra, $\mu:\Sigma \rightarrow [0,1]$ is a normalized measure and $\{T_t\}_{t\in J}$ is a semigroup of preserving measure transformations. For instance, $T_t$ could be the classical Liouville transformation or the corresponding classical transformation associated to the quantum Schr\"{o}dinger transformation. $J$ is usually $\mathbf{R}$ for continuous dynamical systems and $\mathbf{Z}$ for discrete ones.

Let us divide the phase space $\Gamma$ in a partition $Q$ of $m$ small cells $A_{i}$ of measure $\mu (A_{i})$. The entropy of $Q$ is defined as
\begin{equation}\label{entropy partition}
H(Q)=-\sum_{i=1}^{m}\mu(A_{i})\log\mu(A_{i}).
\end{equation}
The KS-entropy of partition $Q$ is given by\footnote{Given two partitions $A$ and $B$ the partition $A\vee B$ is $\{a_i\cup b_j: a_i\in A, b_j\in B\}$. That means $A\vee B$ is a refinement of $A$ and $B$. Given a preserving measure transformation $T_t$, $T^{-j}$ is the inverse of $T_{j}$, i.e. $T^{-j}=T_{j}^{-1}$.}
\begin{equation}\label{KS entropy partition}
h_{\mu}(T,Q)=\lim_{n\rightarrow\infty}\frac{1}{n}H(\vee_{j=0}^{n}T^{-j}Q).
\end{equation}
From this, the KS--entropy $h_{KS}$ of the dynamical system is defined as the supreme of $h_{\mu}(T,Q)$ over all measurable initial partitions of $\Gamma$,
\begin{equation}\label{KS-entropy}
h_{KS}=\sup_{Q}h_{\mu}(T,Q)=\sup_{Q}\{\lim_{n\rightarrow\infty}\frac{1}{n}H(\vee_{j=0}^{n}T^{-j}Q)\}.
\end{equation}
Moreover, from Brudno theorem it can be proved that KS--entropy is the average unpredictability of information of all possible trajectories in the phase space.

On the other hand, it is well--known that chaos in classical dynamics can be defined by the exponential increase of the distance between two trajectories that start from neighboring initial conditions. Quantitatively, it is related with the largest positive Lyapunov exponent of the system \cite{z4}.
The positiveness of largest Lyapunov exponent implies exponential instability of motion. In turn, exponentially unstable motion is chaotic since almost all trajectories are unpredictable in the sense of information theory.

These two quantities, KS--entropy and the Lyapunov exponents, are related to each other by Pesin theorem, which establishes that \cite{LL,P,W}
\begin{equation}\label{pesin theorem}
h_{KS}=\int_{\Gamma}\left[  \sum_{\sigma_{i}(\phi)>0}\sigma
_{i}(\phi)\right]  d^{2(N+1)}\phi ,
\end{equation}
where $\sigma_{i}(\phi)$ are the Lyapunov exponents of the physical system and 2(N+1) is de dimension of the phase space.
When $\sigma$ is constant over all phase space we have
\begin{eqnarray}\label{pesin theorem4}
h_{KS} = \sum_{\sigma>0}\sigma. \nonumber
\end{eqnarray}

At this point, it is appropriate to make a comment on the interest of formula (\ref{pesin theorem}) and its precise physical meaning.
Pesin theorem relates the KS--entropy, i.e. the average unpredictability of information of all possible trajectories in the phase space, with the exponential instability of motion. Then, the main content of Pesin theorem is that $h_{KS}>0$ is a sufficient condition for chaotic motion.

In Section 4, the condition $h_{KS}>0$ will be used to determine chaos in the classical limit of a quantum system, where $h_{KS}$ will be given by a semiclassical condition in the limit $\hbar\rightarrow0$.

\section{Weyl--Wigner--Moyal formalism}
\label{sec:2}
We review the main tools of Wigner transformation for the development of next sections. Given a quantum system we consider its
quantum algebra $\mathcal{A}$. If $\hat{f}\in \mathcal{A}$, then the \textit{Wigner transformation} of $\hat{f}$ is  \cite{Wigner,Symb}
\begin{eqnarray}\label{wigner1}
f(\phi)=\int_{\mathbf{R}^{N+1}}\langle q+\Delta|\,\hat{f}
\,|q-\Delta\rangle e^{2i\frac{p\Delta}{\hbar}}d^{N+1}\Delta, \nonumber
\end{eqnarray}
where $\phi=(q,p)\in\mathbf{R}^{2(N+1)}$,
$p,q,\Delta\in\mathbf{R}^{N+1}$ and $f(\phi)$ is a distribution function over $\mathbf{R}^{2(N+1)}$. From now on we denote $f(\phi)$ by $symb(\hat{f})$.

The set of all distribution functions $\mathcal{A}_{q}=symb(\mathcal{A})$ is called the \textit{quasiclassical} algebra. Given a pure state $\hat{\rho}_{\psi}=|\psi\rangle\langle \psi|$, its Wigner transformation $symb(\hat{\rho}_{\psi})$ can be negative, then the algebra $\mathcal{A}_{q}$ is not classical. For this reason $\rho(\phi)=symb(\hat{\rho})$ is called a quasi-probability distribution, where $\hat{\rho}$ is any density matrix of the quantum system.

Given two operators $\hat{f}, \hat{g} \in \mathcal{A}$, we can also introduce the \textit{star product }\cite{Bayern}
\begin{eqnarray}\label{wigner2}
symb(\hat{f} \ \hat{g})=symb(\hat{f})\ast
symb(\hat{g})=(f \ast  g)(\phi)=
f(\phi)\exp\left( -\frac{i\hbar}{2}\overleftarrow{\partial}_{a}
\omega^{ab}\overrightarrow{\partial}_{b}\right)  g(\phi), \nonumber
\end{eqnarray}
where $f(\phi)=sym(\hat{f})$, $g(\phi)=sym(\hat{g})$ and $\omega^{ab}$ is the metric tensor of the phase space $\Gamma$.

The \textit{Moyal bracket} is the symbol corresponding to the quantum commutator, i.e.
\begin{eqnarray}\label{wigner3}
\{f,g\}_{MB}=\frac{1}{i\hbar}(f\ast g-g\ast f)=symb\left(\frac{1}{i\hbar
}[\hat{f},\hat{g}]\right).\nonumber
\end{eqnarray}
It can be proved that
\begin{eqnarray}\label{wigner4}
(f\ast g)(\phi)=f(\phi)g(\phi)+\mathcal{O}(\hbar) \,\,\,\,\,\,\,\,\ and \,\,\,\,\,\,\,\,\,
\{f,g\}_{MB}=\{f,g\}_{PB}+\mathcal{O}(\hbar^{2}),
\end{eqnarray}
where
\begin{eqnarray}\label{wigner5}
\{f,g\}_{PB}=\sum_{i=1}^{N+1}\left(\frac{\partial f}{\partial q_i}\frac{\partial g}{\partial p_i}-\frac{\partial f}{\partial p_i}\frac{\partial g}{\partial q_i}\right)\nonumber
\end{eqnarray}
is the \emph{Poisson bracket}.

The transformations $symb$ and $symb^{-1}$
define an isomorphism between the quantum algebra $\mathcal{A}$ and
the quasiclassical algebra of
distribution functions $\mathcal{A}_{q}$
\begin{eqnarray}\label{wigner6}
symb:\mathcal{A}\rightarrow \mathcal{A}_{q} \,\,\,\,\,\,\,\,\ , \,\,\,\,\,\,\,\,\,
symb^{-1}:\mathcal{A}_{q}\rightarrow\mathcal{A}. \nonumber
\end{eqnarray}
The mapping so defined is the \textit{Weyl--Wigner--Moyal symbol}. When
$\hbar\rightarrow0$, $\mathcal{A}_{q}$ tends to $\mathcal{A}_{cl}$, where
$\mathcal{A}_{cl}$ is the classical algebra of observables\footnote{By ``$\mathcal{A}_{q}$ tends to $\mathcal{A}_{cl}$" we mean that in the classical limit, $\hbar\rightarrow0$, the quassiclassical algebra $\mathcal{A}_{q}$ tends to the commutative algebra of functions defined over $\Gamma$, i.e. $\mathcal{A}_{cl}$, where $\hbar$ is the parameter of deformation quantization.}.

A relevant property of the Wigner transformation is \cite{Wigner}
\begin{eqnarray}\label{wigner8}
&\langle\hat{O}\rangle_{\hat{\rho}}=(\hat{\rho}|\hat{O})=\langle symb(\hat{\rho}), symb(\hat{O})\rangle=
\langle\rho(\phi), O(\phi)\rangle=\nonumber \\
&\int_{\mathbf{R}^{2(N+1)}} d^{2(N+1)}\phi\rho(\phi)\,O(\phi),
\end{eqnarray}
where $\langle f,g\rangle$ is the scalar product between $f$ and $g$, and $(\hat{\rho}|\hat
{O})$ is a notation for the mean value of $\hat{O}$ in $\hat{\rho}$. In other words, $(\hat{\rho}|\hat
{O})=\langle\hat{O}\rangle_{\hat{\rho}}=tr(\hat{\rho} \hat{O})$ is the action of the functional $\hat{\rho}$ on the observable $\hat{O}$. Let us make a brief remark about formula (\ref{wigner8}). It says that the mean value of an observable $\hat{O}$ in a state $\hat{\rho}$
can be calculated, equivalently, in the quantum algebra $\mathcal{A}$ or in the quasiclassical algebra $\mathcal{A}_q$, i.e as the trace of $\hat{\rho} \hat{O}$ or as the scalar product $\int_{\mathbf{R}^{2(N+1)}} d^{2(N+1)}\phi\rho(\phi)\,O(\phi)$.

A particular case of Eq. (\ref{wigner8}) is when $\hat{O}$ is the identity $\hat{I}$,
\begin{eqnarray}\label{wigner9}
\langle\hat{I}\rangle_{\hat{\rho}}=(\hat{\rho}|\hat
{I})=\langle symb(\hat{\rho}),symb(\hat{I})\rangle=\langle\rho(\phi),\,I(\phi)\rangle=\int_{\mathbf{R}^{2(N+1)}} d^{2(N+1)}\phi\rho(\phi)=1, \nonumber
\end{eqnarray}
which is nothing but the normalization condition for the state $\hat{\rho}$.

In next section we use the Wigner transformation property given by Eq. (\ref{wigner8}) to express Pesin theorem by means of quantum mean values.

\section{Pesin theorem expressed in terms of quantum mean values: A semiclassical condition for chaos}

With the mathematical background of previous section and the definitions of Section 2 we will write the Pesin theorem in terms of quantum mean values.

As a starting point we make the following assumptions. Let $S$ be a quantum system with its quantum algebra $\mathcal{A}$. We assume $S$ has a classical limit $S_{cl}$\footnote{The classical algebra $\mathcal{A}_{cl}$ of $S_{cl}$ is the limit of the quasiclassical $\mathcal{A}_{q}$ of $S$ when $\hbar\rightarrow0$, i.e. $lim_{\hbar\rightarrow0}\mathcal{A}_q =\mathcal{A}_{cl}$.}, which is a dynamical system with a phase space $\Gamma$ and a classical group of transformations $\{T_t\}$.
Since the process of generating the KS--entropy involves a discrete sequence of steps, the quantum evolution of $S$ is forced to be discretized\footnote{For instance, discretized evolutions are used in Hamiltonians with a time-dependent potential. In such cases, it is common to take $\hat{U}(n)=\hat{F}(\tau n)$, where $\hat{F}$ is the Floquet operator and $\tau$ is the periodicity of the potential.}. We consider $\hat{U}(j)=e^{-i\frac{\hat{H}}{\hbar}\alpha j}$ as the discretized evolution operator\footnote{In an irreversible process the effective Hamiltonian of the system describes the system in interaction with its environment. In general, it is not a self-adjoint operator, $\hat{H}\neq\hat{H}^{\dag}$.} associated with the classical transformation $T_j$, where $T$ is taken as $T_1$ and the real parameter $\alpha$ defines the time steps.

Then, we show a property that is the key point to express Pesin theorem by means of quantum mean values. Given a partition $Q=\{A_1,...,A_m\}$ of $\Gamma$, we write the measure of an element $A_i$ of $Q$ at time $t$ as
the trace of an appropriate operator $\hat{I}_{A_i}$ at time $t$.
More precisely, let $I_{A_i(t)}(\phi)$ be the characteristic function of $A_i(t)$, where $A_i(t)=T_t(A_i)$ and $A_i(t)$ is $A_i$ at time $t$.
Then, by definition we have
\begin{eqnarray}\label{pesin1}
&\mu(A_{i}(t))=\int d^{2(N+1)}\phi I_{A_{i}(t)}(\phi)=\langle
I_{A_{i}(t)}(\phi),\,I(\phi)\rangle=\nonumber \\
&\langle symb(\hat{I}_{A_i}(t)),symb(\hat{I})\rangle=(\hat{I}_{A_i}(t)|\hat{I})=\langle\hat{I}\rangle_{\hat{I}_{A_i}(t)},
\end{eqnarray}
where we have used the Wigner transformation property (see Eq. (\ref{wigner8})).

Therefore, $\mu(A_{i}(t))=(\hat{I}_{A_i}(t)|\hat{I})$, which means that the measure of $A_{i}$ at time $t$ is equal to the trace of the operator $\hat{I}_{A_i}$ at time $t$, where $\hat{I}_{A_i}(t)$ is the Wigner transformation of the characteristic function of $A_i(t)$.

Next step is to write a semiclassical version ($\hbar\approx0$) of the KS--entropy of any partition using the formula (\ref{pesin1}). Consider a partition $Q=\{A_1,...,A_m\}$ of $\Gamma$. Then, we have the partition $B(-n)=\vee_{j=0}^{n}T^{-j}Q$.
Let $B(k_0, k_1, ...,k_n)=\bigcap_{j=0}^n T^{-j} A_{k_j}$ be an element of $B(-n)$. Using Eq. (\ref{ap2}) of Appendix A we can give an expression for $\mu(B(k_0, k_1, ...,k_n))$. We have
\begin{eqnarray}\label{pesin2}
\mu(B(k_0, k_1, ...,k_n))=(\prod_{j=0}^n\hat{I}_{A_{k_j}}(j)|\hat{I}) \,\,\,\,\,\,\ when \,\,\,\ \hbar\approx0,
\end{eqnarray}
where $\hat{I}_{A_{k_j}}(j)=\hat{U}(j)\hat{I}_{A_{k_j}}(0)\hat{U}(j)^{\dag}$ is $\hat{I}_{A_{k_j}}(0)$ after $j$ steps.

Therefore, if we replace $\mu(B(k_0, k_1, ...,k_n))$ by $(\prod_{j=0}^n\hat{I}_{A_{k_j}}(j)|\hat{I})$ in Eq. (\ref{KS entropy partition}), we obtain
\begin{eqnarray}\label{pesin3}
&h_{\mu}(T,Q)=\lim_{n\rightarrow\infty}\frac{1}{n}H(B(-n))=\nonumber \\
&-\lim_{n\rightarrow\infty}\frac{1}{n}\sum_{(k_0, k_1, ...,k_n)}^{R_n}\mu(B(k_0, k_1, ...,k_n))\log\mu(B(k_0, k_1, ...,k_n))=\nonumber \\
&-\lim_{n\rightarrow\infty}\frac{1}{n}\sum_{(k_0, k_1, ...,k_n)}^{R_n}(\prod_{j=0}^n\hat{I}_{A_{k_j}}(j)|\hat{I})\log(\prod_{j=0}^n\hat{I}_{A_{k_j}}(j)|\hat{I}), \nonumber
\end{eqnarray}
where $R_n$ is the number of elements of $B(-n)$\footnote{Indeed, $R_n$ is well known as the topological entropy of $B(-n)$. Roughly speaking, $R_n$ ``measures" the degree of mixing of a dynamical system as it evolves in time. Typically, in a fully chaotic system the formation of fractal structures in a chaotic sea can produce numerous sets $B(k_0, k_1, ...,k_n)$ and therefore an increasing of $R_n$.}.

Then, from Eqns. (\ref{KS-entropy}) and (\ref{pesin theorem}), we obtain the Pesin theorem in terms of quantum mean values
\begin{eqnarray}\label{quantum pesin theorem}
&\sup_{Q}\{-\lim_{n\rightarrow\infty}\frac{1}{n}\sum_{(k_0, k_1, ...,k_n)}^{R_n}(\prod_{j=0}^n\hat{I}_{A_{k_j}}(j)|\hat{I})\log(\prod_{j=0}^n\hat{I}_{A_{k_j}}(j)|\hat{I})\} \nonumber \\
&=\int_{\Gamma}\left[  \sum_{\sigma_{i}(\phi)>0}\sigma
_{i}(\phi)\right]  d^{2(N+1)}\phi  \,\,\,\,\,\,\ \,\,\,\,\,\,\ when \,\,\,\ \hbar\approx0.
\end{eqnarray}

Formula (\ref{quantum pesin theorem}) implies that if we have a quantum system $S$, with a classical limit $S_{cl}$, the positive Lyapunov exponents of $S_{cl}$ are related with the supreme of an expression which involves the mean values $(\prod_{j=0}^n\hat{I}_{A_{k_j}}(j)|\hat{I})$. Moreover, it gives an alternative method for calculating Lyapunov exponents of the classical limit of a quantum system.

As the number $R_n$ is usually hard to calculate,  usefulness of Eq. (\ref{quantum pesin theorem}) seems to be restricted to simple cases where $R_n$ is trivial.\footnote{For instance, if $R_n$ is bounded for all $n$, then from Eq. (\ref{KS-entropy}) it follows that $\sup_{Q}\{...\}=0$. From Eq. (\ref{pesin theorem}) we obtain $\int_{\Gamma}\left[  \sum_{\sigma_{i}(\phi)>0}\sigma
_{i}(\phi)\right]  d^{2(N+1)}\phi = 0$, which implies that $\sigma_{i}(\phi)=0$ for all $i$. Therefore, in such case the system is not chaotic.}
However, if we are only interested in knowing if $S_{cl}$ is chaotic or not, we do not need to perform the supreme of Eq. (\ref{quantum pesin theorem}) explicitly.
Instead, with the help of the following lemma, it is enough to focus in the asymptotic behavior ($n\rightarrow \infty$) of $\mu(B(k_0, k_1, ...,k_n))$ to ensure the existence of positive Lyapunov exponents and to conclude that $S_{cl}$ is chaotic. The lemma states \cite{LL}
\begin{eqnarray}\label{lemma}
\mu(B(k_0, k_1, ...,k_n)) \,\ decreases \,\ exponentially  \Longrightarrow KS-entropy >0.
\end{eqnarray}

This lemma is a sufficient condition for chaos. It says that chaos is governed by the exponential decay of $\mu(B(k_0, k_1, ...,k_n))$ in the limit $n\rightarrow\infty$. Physically, this asymptotic limit means looking at the system for large times, without taking into account the details of the chaotic dynamics at finite times, like the formation of fractal structures in a chaotic sea or the folding of trajectories.
Taking into account that $\mu(B(k_0, k_1, ...,k_n)) =(\prod_{j=0}^n\hat{I}_{A_{k_j}}(j)|\hat{I})$ when $\hbar\approx0$ (see Eq. \ref{pesin2}), then lemma of Eq. (\ref{lemma}) becomes
\begin{eqnarray}\label{semilemma}
(\prod_{j=0}^n\hat{I}_{A_{k_j}}(j)|\hat{I}) \,\ decreases \,\ exponentially  \Longrightarrow KS-entropy >0,
\end{eqnarray}
which provides a condition for chaos in the classical limit $S_{cl}$.

Summing up the previous steps up to Eq. (\ref{semilemma}), we can obtain a method to determine if $S_{cl}$ is chaotic or not. The prescription of the method is as follows:\begin{itemize}

\item[$(a)$] Take a generic partition $Q=\{A_i:i=1,...,m\}$ of phase space $\Gamma$ of $S_{cl}$.\\

\item[$(b)$] For any n--tuple $(k_0,k_1,...,k_n)$ with $k_j\in\{1,...,m\}$ calculate the operators $\hat{I}_{A_{k_j}}(j) =\hat{U}(j)\hat{I}_{A_{k_j}}(0)\hat{U}(j)^{\dag}$, where $\hat{I}_{A_{k_j}}(0)=sym^{-1}(I_{A_{k_j}}(\phi))$.\\

\item[$(c)$] Then, perform $(\prod_{j=0}^n\hat{I}_{A_{k_j}}(j)|\hat{I})$ for all $n$. \\

\item[$(d)$] Finally, if $(\prod_{j=0}^n\hat{I}_{A_{k_j}}(j)|\hat{I})$ decreases exponentially when $n\rightarrow\infty$, then KS-entropy of $S_{cl}$ is positive. Therefore, $S_{cl}$ is chaotic.
\end{itemize}

In next section we see how prescription $(a)-(d)$ works with an example.

\section{Physical relevance}

In order to illustrate the physical relevance of the condition given by Eq. (\ref{semilemma}), we apply the prescription $(a)-(d)$ to an example of the decoherence literature: a phenomenological Gamow model type \cite{Omnes1,PRE}.
This model consists of a single oscillator embedded in an environment composed of a large bath of noninteracting oscillators, which can be considered as a continuum.

The degeneration of this system prevents the application of perturbation theory. Instead, we can apply an analytical extension of the Hamiltonian \cite{PRE,antoniou,Gadella,letterpolos,Ordonito-dec,4A} to obtain an non-hermitian effective Hamiltonian $\hat{H}_{eff}$. Non-hermiticity of $H_{eff}$ yields two set of eigenvectors $\{\langle \widetilde{m}|\}_{m=0}^{\infty}$ and $\{|n\rangle\}_{n=0}^{\infty}$ (left and right eigenvectors, respectively), which satisfy \cite{moiseyev}
\begin{eqnarray}\label{gamow1}
\hat{H}_{eff}|n\rangle=z_n |n\rangle , \,\,\,\,\,\,\,\,\,\, \langle \widetilde{n}|H_{eff}=\langle \widetilde{n}|z_j , \,\,\,\,\,\,\,\,\,\,   n\in\mathbf{N}_0, \nonumber
\end{eqnarray}
\begin{eqnarray}\label{gamow2}
\,\,\,\,\,\,\,\,\,\,\,\ \langle \widetilde{m}|n\rangle=\delta_{mn} \,\,\,\,\,\,\,\,\,\, (bi-orthogonality), \nonumber
\end{eqnarray}
\begin{eqnarray}\label{gamow3}
\sum_{n=0}^{\infty} |n\rangle \langle\widetilde{n}|=\hat{I} \,\,\,\,\,\,\,\,\,\, (completeness).\nonumber
\end{eqnarray}

The effective Hamiltonian $\hat{H}_{eff}$ is given by
\begin{eqnarray}\label{gamow3}
\hat{H}_{eff}=\sum_{n=0}^{\infty}z_{n}|n\rangle
\langle\widetilde{n}|,\nonumber
\end{eqnarray}
where $z_{n}=n(\omega_0-i\gamma_0)$ are complex eigenvalues, except $z_0=\omega_0$,  $\gamma_0$ is associated with the decoherence time $t_R=\frac{\hbar}{\gamma_0}$ and $\omega_0$ is the natural frequency of the single oscillator \cite{Omnes1}.

From formula (\ref{ap5}) of Appendix B, we can expand the operators $\hat{I}_{A_{k_j}}(j)$ in the bi-orthogonal basis $\{|r\rangle \langle\widetilde{s}|\}_{r,s\in\mathbf{N}_0}$
\begin{eqnarray}\label{gamow4}
&\hat{I}_{A_{k_j}}(j)=\alpha_{A_{k_j}}(0,0)|0\rangle\langle
0|+\sum_{r=1}^{\infty}\alpha_{A_{k_j}}(r,r)e^{-2\frac{\gamma_0}{\hbar}r\alpha j}|r\rangle\langle\widetilde{r}|+ \nonumber\\
&+\sum_{s=1}^{\infty}\alpha_{A_{k_j}}(0,s)e^{i\frac{\omega_0}{\hbar}(s-1)\alpha j}e^{-\frac{\gamma_0}{\hbar}s\alpha j}|0\rangle\langle\widetilde{s}| +\nonumber \\
&+\sum_{r=1}^{\infty}\alpha_{A_{k_j}}(r,0)e^{-i\frac{\omega_0}{\hbar}(r-1)\alpha j}e^{-\frac{\gamma_0}{\hbar}r\alpha j}|r\rangle\langle0|+ \nonumber \\
&+\sum_{r,s>0,r\neq s}^{\infty}\alpha_{A_{k_j}}(r,s)e^{-\frac{\gamma_0}{\hbar}(r+s)\alpha j}|r\rangle\langle\widetilde{s}|.
\end{eqnarray}

From Eq. (\ref{gamow4}), we see that for $j\gg \frac{\hbar}{\alpha \gamma_0}=\frac{t_R}{\alpha}$ all the sums decay exponentially. Then, we can neglect these terms and obtain
\begin{eqnarray}\label{gamow5}
\hat{I}_{A_{k_j}}(j)\simeq \alpha_{A_{k_j}}(0,0)|0\rangle\langle 0| \,\,\,\,\,\,\,\,\, for \,\ all \,\ j\gg \frac{t_R}{\alpha} \,\ with \,\ j=1,...,n.
\end{eqnarray}
The coefficient $\frac{t_R}{\alpha}$ can be interpreted as an adimensional relaxation time, where parameter $\alpha$ defines the time steps of the discretized evolution.

From Eq. (\ref{gamow5}), we can obtain an asymptotic expression for $\prod_{j=0}^{n}\hat{I}_{A_{k_j}}(j)$, when $n\gg \frac{t_R}{\alpha}$\footnote{From Eq. (\ref{gamow4}), it follows that if $n\gg \frac{t_R}{\alpha}$, then $\hat{I}_{A_{k_n}}(n)\simeq \alpha_{A_{k_n}}(0,0)|0\rangle\langle 0|$ is diagonal. Thus, $\prod_{j=0}^{n}\hat{I}_{A_{k_j}}(j)=\hat{I}_{A_{k_0}}(0).\hat{I}_{A_{k_1}}(1)...\hat{I}_{A_{k_n}}(n)
\simeq\hat{I}_{A_{k_0}}(0).\hat{I}_{A_{k_1}}(1)...\alpha_{A_{k_n}}(0,0)|0\rangle\langle 0|=\left(\prod_{j=0}^{n}\alpha_{A_{k_j}}(0,0)\right)|0\rangle\langle 0|$ is diagonal, regardless if operators $\hat{I}_{A_{k_0}}(0), \hat{I}_{A_{k_1}}(1),...,\hat{I}_{A_{k_{n-1}}}(n-1)$ are diagonals or not.},
\begin{eqnarray}\label{gamow6}
\prod_{j=0}^{n}\hat{I}_{A_{k_j}}(j)\simeq \left(\prod_{j=0}^{n}\alpha_{A_{k_j}}(0,0)\right)|0\rangle\langle 0| \,\,\,\,\,\,\,\,\, for \,\ n\gg \frac{t_R}{\alpha},
\end{eqnarray}
and therefore,
\begin{eqnarray}\label{gamow7}
(\prod_{j=0}^{n}\hat{I}_{A_{k_j}}(j)|\hat{I})\simeq \prod_{j=0}^{n}\alpha_{A_{k_j}}(0,0) \,\,\,\,\,\,\,\,\, for \,\ n\gg \frac{t_R}{\alpha}.
\end{eqnarray}

Up to Eq. (\ref{gamow7}) we have completed the steps $(a)-(c)$ of our prescription.
The last step is to check that $(\prod_{j=0}^n\hat{I}_{A_{k_j}}(j)|\hat{I})$ decays exponentially when $n\rightarrow\infty$.

First, we note that when $j\longrightarrow \infty$ we have
\begin{eqnarray}\label{gamow8}
&\mu(A_{k_j}(j))=(\hat{I}_{A_{k_j}}(j)|\hat{I})=\alpha_{A_{k_j}}(0,0)+\nonumber \\
&+\sum_{n=1}^{\infty}\alpha_{A_{k_j}}(n,n)e^{-2\frac{\gamma_0}{\hbar}n\alpha j}\longrightarrow \alpha_{A_{k_j}}(0,0).
\end{eqnarray}
Since we consider classical motion is bounded\footnote{Typically, phase space of a non-integrable chaotic system is a compact manifold. If motion is regular and integrable, the phase space can be taken as a torus.}, we can consider phase space $\Gamma$ is normalized. Then, from Eq. (\ref{gamow8}), we have
\begin{eqnarray}\label{gamow9}
\mu(\Gamma)=1>\mu(A_{k_j}(j))\longrightarrow\alpha_{A_{k_j}}(0,0).
\end{eqnarray}
Also, given that $\mu(A_{k_j}(j))\geq0$ for all $j$, it follows that $0\leq\alpha_{A_{k_j}}(0,0)<1$.

Moreover, if $\mu(A_{k_{j_0}}(j_0))=0$ for some $k_{j_0}$, since $\mu(A_{k_{j_0}}(j_0))=(\prod_{j=0}^{n}\hat{I}_{A_{k_{j_0}}}(j_0)|\hat{I})$, then $(\prod_{j=0}^{n}\hat{I}_{A_{k_{j_0}}}(j_0)|\hat{I})=0$. Therefore,
\begin{equation}
(\prod_{j=0}^{n}\hat{I}_{A_{k_{j_0}}}(j_0)|\hat{I})\log(\prod_{j=0}^{n}\hat{I}_{A_{k_{j_0}}}(j_0)|\hat{I})= 0, ~\footnote{If $f(x)=x\log(x)$, then by definition  $f(0)=0$.}
\end{equation}
and it does not contribute to the semiclassical version of KS--entropy of Eq. (\ref{quantum pesin theorem}).
This means that we can consider $\mu(A_{k_j}(j))>0$ for all $j$. Thus, we have
\begin{eqnarray}\label{gamow10}
0<\alpha_{A_{k_j}}(0,0)<1 \,\,\,\,\,\,\,\,\, for \,\ all \,\ j=1,...,n.
\end{eqnarray}

If we call $\delta_1=min\{\alpha_{A_{k_j}}(0,0):k_j=1,...,m\}$ and $\delta_2=max\{\alpha_{A_{k_j}}(0,0):k_j=1,...,m\}$, then from Eq. (\ref{gamow10}) we have
\begin{eqnarray}\label{gamow11}
\delta_1^{n+1}<\prod_{j=0}^{n}\alpha_{A_{k_j}}(0,0)<\delta_2^{n+1}.
\end{eqnarray}
Finally, from Eqns. (\ref{gamow7}) and (\ref{gamow11}), we obtain
\begin{eqnarray}\label{gamow12}
\delta_1^{n+1}<(\prod_{j=0}^{n}\hat{I}_{A_{k_j}}(j)|\hat{I})<\delta_2^{n+1}\,\,\,\,\,\,\,\,\, for \,\ n\gg \frac{t_R}{\alpha}.
\end{eqnarray}

Eq. (\ref{gamow12}) implies that $(\prod_{j=0}^{n}\hat{I}_{A_{k_j}}(j)|\hat{I})$ decreases exponentially. Therefore, by prescription $(a)-(d)$, we conclude the positiveness of Lyapunov exponents of classical limit of the phenomenological Gamow model.  Then, its classical limit is chaotic.

\section{Conclusions}

In this paper we used properties of Wigner transformation in order to express classical quantities by means of quantum mean values. In particular, we translated the quantities involved in Pesin theorem and we obtained a version of Pesin theorem expressed in terms of quantum mean values, which relates the Lyapunov exponents of the classical limit of a system with the mean value of the projectors that correspond to  characteristic functions on phase space.

Moreover, from the modified version of Pesin theorem, we obtained a method (the prescription $(a)-(d)$ of Section 4) to determine if the classical limit of a quantum system is chaotic or not.
The core of this method is the step $(d)$, which establishes that if we have a quantum system where $(\prod_{j=0}^n\hat{I}_{A_{k_j}}(j)|\hat{I})$ exponentially decays when $n\rightarrow\infty$, then the KS-entropy of its classical limit $S_{cl}$ is positive.
This also implies that the system must have positive Lyapunov exponents and, therefore, its classical limit must be chaotic. Summing up,
\begin{eqnarray}
(\prod_{j=0}^n\hat{I}_{A_{k_j}}(j)|\hat{I}) \,\ exponentially \,\ decreases  \Longrightarrow S_{cl} \,\ is \,\ chaotic. \nonumber
\end{eqnarray}

Finally, in Section 5, we applied our method to a phenomenological Gamow-type model and we concluded that its classical limit is chaotic. The exponential decay of $(\prod_{j=0}^n\hat{I}_{A_{k_j}}(j)|\hat{I})$ occurs for $n>> \frac{t_R}{\alpha}$, where $t_R$ is the   decoherence time and $\alpha$ the real parameter which defines the time steps.

Furthermore, from Pesin theorem expressed in terms of quantum mean values, the quantity $(\prod_{j=0}^n\hat{I}_{A_{k_j}}(j)|\hat{I})$ can be related with the positive Lyapunov exponents.
This suggests that decoherence time of Gamow-type model could be related with positive Lyapunov exponents of its classical limit. Here we see an interesting hypothesis about a possible relationship between decoherence time and Lyapunov exponents and we hope it will be corroborated in future researches with more examples and theoretical essays.


\section*{Acknowledgments}
This paper was supported partially by the CONICET (Argentine research Council), the IFIR (Instituto de F\'{i}sica de Rosario), the IFLP (Instituto de F\'{i}sica de La Plata) and the Buenos Aires University.

The authors would like to acknowledge the anonymous reviewer for helpful comments on the original manuscript.

\appendix
\clearpage 
\addappheadtotoc
\appendixpage

\section{The classical quantity $\mu (B(k_0,k_1,...,k_n))$ expressed as a quantum mean value}

In order to evaluate the KS entropy, we have to generate the following partition
\begin{eqnarray}\label{ap1}
B(-n)= \bigvee_{j=0}^n T^{-j} Q = \{ \bigcap_{j=0}^n T^{-j} A_{k_j} : A_{k_j} \in Q \}, \nonumber
\end{eqnarray}

If $B(k_0,k_1,...,k_n)=\bigcap_{j=0}^n T^{-j} A_{k_j}$ is a generic element of $B(-n)$, then the measure of $B(k_0,k_1,...,k_n)$ is
\begin{eqnarray}\label{ap2}
&\mu(B(k_0,k_1,...,k_n))=\mu(\bigcap_{j=0}^n T^{-j} A_{k_j})=\int_{\bigcap_{j=0}^n T^{-j} A_{k_j}}d^{2(N+1)}\phi=
\int_{\Gamma} I_{\bigcap_{j=0}^n T^{-j} A_{k_j}}(\phi) d^{2(N+1)}\phi\nonumber\\
&=\int_{\Gamma}\prod_{j=0}^n I_{A_{k_j}}(T^j\phi) d^{2(N+1)}\phi=
\langle\prod_{j=0}^nI_{A_{k_j}}\circ T^j(\phi),I(\phi)\rangle=
\langle symb(\prod_{j=0}^n\widehat{I_{A_{k_j}}\circ T^j}),symb(\hat{I})\rangle\nonumber \\
&=(\prod_{j=0}^n\widehat{I_{A_{k_j}}\circ T^j}|\hat{I})=(\prod_{j=0}^n\hat{I}_{A_{k_j}}(j) |\hat{I}),
\end{eqnarray}
\noindent where we have used the following properties:

\begin{itemize}
\item[$\bullet$]
The characteristic function of an intersection of sets is the product of the characteristic functions of each set.

~

\item[$\bullet$] If $T$ is bijective, then $I_{T^{-j}A_{k_j}}(\phi)= I_{A_{k_j}}(T^j\phi)$.

~

\item[$\bullet$] If $\hbar\approx 0$, then $symb(\prod_j^{n}\hat{f}_j)(\phi)\simeq\prod_j ^{n}f_j(\phi)$, where we have neglected terms of order $\mathcal{O}(\hbar)$). This property is the generalization of Eq. (\ref{wigner4}) for a product of $n$ functions $f_i$.

~

\item[$\bullet$] $\widehat{I_{A_{k_j}}\circ T^j}=\hat{I}_{A_{k_j}}(j)=\hat{U}(j)\hat{I}_{A_{k_j}}(0)\hat{U}(j)^{\dag}$, where $\hat{U}(j)=e^{-\frac{i}{\hbar}\hat{H}\alpha j}$ is the evolution operator and $\alpha$ is a real parameter which defines the time steps. This property is a consequence of the formula (\ref{wigner8}).
\end{itemize}

\section{An expansion for  operators $\hat{I}_{A_{k_j}}$}
We consider a Hamiltonian of the form
\begin{eqnarray}\label{ap3}
\hat{H}=\sum_r z_r \vert r\rangle\langle\widetilde{r}|, \nonumber
\end{eqnarray}
where $z_r=Re(z_r)+iIm(z_r)$ are complex eigenvalues and $\{\vert r\rangle\},\{\langle \widetilde{s}|\}$ are its two sets of eigenvectors, left and righ respectively \cite{moiseyev}. Then we have
\begin{eqnarray}\label{ap4}
\hat{I}_{A_{k_j}}(0)=\sum_{r,s}\alpha_{A_{k_j}}(r,s)\vert r\rangle\langle \widetilde{s}|. \nonumber
\end{eqnarray}

Therefore,
\begin{eqnarray}\label{ap5}
&\hat{I}_{A_{k_j}}(j)=e^{-\frac{i}{\hbar}\hat{H} \alpha j} \left(\sum_{r,s}\alpha_{A_{k_j}}(r,s) \vert r \rangle\langle \tilde {s} \vert\right)e^{\frac{i}{\hbar}\hat{H}^{\dag} \alpha j}=\nonumber \\
&e^{-(\frac{i}{\hbar} \sum_p z_p \vert p \rangle\langle \widetilde{p}|)\alpha j}\left(\sum_{r,s}\alpha_{A_{k_j}}(r,s) \vert r \rangle\langle \widetilde{s}| \right) e^{(\frac{i}{\hbar}\sum_q z_q^* \vert q \rangle\langle \widetilde{q}|)\alpha j} \nonumber \\
&=\sum_p\sum_q\alpha_{A_{k_j}}(p,q)e^{(-\frac{i}{\hbar}z_p)\alpha j}e^{(\frac{i}{\hbar}z_q^*)\alpha j}\vert p\rangle\langle\widetilde{q}|,
\end{eqnarray}
where we have used the exponential of an operator ($e^{\hat{A}}=\sum_{k=0}^\infty \frac{\hat{A}^k}{k!}$) and the orthogonal relations of the projectors $\vert r \rangle\langle \widetilde {s} \vert$, that is
\begin{eqnarray}\label{ap6}
&(\vert r \rangle\langle \widetilde {r} \vert)^{k}=\vert r \rangle\langle \widetilde {r} \vert , \,\ and \nonumber \\
&\langle \widetilde {s} |r\rangle=0 \,\,\,\,\, if \,\,\,\,\, r\neq s.
\end{eqnarray}

\end{document}